\documentclass[aps,prd,preprint,floatfix,nofootinbib,superscriptaddress]{revtex4}

\usepackage{graphicx,subfigure,epsfig,epsf,color}
\usepackage{amssymb,amsmath}
\usepackage{slashed}
\usepackage{feynmf}
\usepackage{array}
\usepackage{subcaption}
\usepackage{hyperref}

\newcolumntype{C}[1]{>{\centering\let\newline\\\arraybackslash\hspace{0pt}}m{#1}}

\newcommand{\beq}{\begin{equation}}
\newcommand{\eeq}{\end{equation}}
\newcommand{\bea}{\begin{eqnarray}}
\newcommand{\eea}{\end{eqnarray}}
\newcommand{\beas}{\begin{eqnarray*}}
\newcommand{\eeas}{\end{eqnarray*}}
\newcommand{\bcr}{\begin{center}}

\def\Re{{\cal R \mskip-4mu \lower.1ex \hbox{\it e}\,}}
\def\Im{{\cal I \mskip-5mu \lower.1ex \hbox{\it m}\,}}

\def\tev{\,{\ifmmode\mathrm {TeV}\else TeV\fi}}
\def\gev{\,{\ifmmode\mathrm {GeV}\else GeV\fi}}
\def\mev{\,{\ifmmode\mathrm {MeV}\else MeV\fi}}
\def\to{\rightarrow}

\begin{document}

\def\issue(#1,#2,#3){#1 (#3) #2} 
\def\APP(#1,#2,#3){Acta Phys.\ Polon.\ \issue(#1,#2,#3)}
\def\ARNPS(#1,#2,#3){Ann.\ Rev.\ Nucl.\ Part.\ Sci.\ \issue(#1,#2,#3)}
\def\CPC(#1,#2,#3){comp.\ Phys.\ comm.\ \issue(#1,#2,#3)}
\def\CIP(#1,#2,#3){comput.\ Phys.\ \issue(#1,#2,#3)}
\def\EPJC(#1,#2,#3){Eur.\ Phys.\ J.\ C\ \issue(#1,#2,#3)}
\def\EPJD(#1,#2,#3){Eur.\ Phys.\ J. Direct\ C\ \issue(#1,#2,#3)}
\def\IEEETNS(#1,#2,#3){IEEE Trans.\ Nucl.\ Sci.\ \issue(#1,#2,#3)}
\def\IJMP(#1,#2,#3){Int.\ J.\ Mod.\ Phys. \issue(#1,#2,#3)}
\def\JHEP(#1,#2,#3){J.\ High Energy Physics \issue(#1,#2,#3)}
\def\JPG(#1,#2,#3){J.\ Phys.\ G \issue(#1,#2,#3)}
\def\MPL(#1,#2,#3){Mod.\ Phys.\ Lett.\ \issue(#1,#2,#3)}
\def\NP(#1,#2,#3){Nucl.\ Phys.\ \issue(#1,#2,#3)}
\def\NIM(#1,#2,#3){Nucl.\ Instrum.\ Meth.\ \issue(#1,#2,#3)}
\def\PL(#1,#2,#3){Phys.\ Lett.\ \issue(#1,#2,#3)}
\def\PRD(#1,#2,#3){Phys.\ Rev.\ D \issue(#1,#2,#3)}
\def\PRL(#1,#2,#3){Phys.\ Rev.\ Lett.\ \issue(#1,#2,#3)}
\def\PTP(#1,#2,#3){Progs.\ Theo.\ Phys. \ \issue(#1,#2,#3)}
\def\RMP(#1,#2,#3){Rev.\ Mod.\ Phys.\ \issue(#1,#2,#3)}
\def\SJNP(#1,#2,#3){Sov.\ J. Nucl.\ Phys.\ \issue(#1,#2,#3)}

\bibliographystyle{revtex}

\title{Blackbody Radiation in $q$-deformed Statistics}

\author{Atanu~Guha}
\email[]{p2014401@goa.bits-pilani.ac.in}
\author{Prasanta~Kumar~Das}
\email[]{Author(corresponding):pdas@goa.bits-pilani.ac.in}
\affiliation{Birla Institute of Technology and Science-Pilani, Department of Physics, 
Goa campus, NH-17B, Zuarinagar, Goa-403726, India }

\date{\today}

\begin{abstract} 
 More general canonical ensemble which gives rise the generalized statistics or $q$-deformed 
 statistics can represent the realistic scenario than the ideal one,  with proper parameter sets involved. 
 We study the Planck's law of blackbody radiation, Wein's and Rayleigh-Jeans radiation
 formulae from the point of view of $q$-deformed statistics.  We find that the blackbody 
 energy spectrum curve for a given temperature $T$ corresponding to different $q$ values differs 
 from  each other: the location of the peak(i.e. $\nu_m$) of the energy distribution $u_\nu$ (corresponding to different $q$ )
 shifted towards higher $\nu$ for higher $q$. From the $q$-deformed Wein's displacement law, 
 we find that $\lambda_m T$ varies from $0.0029~\rm{m~K}$ to $0.0017~\rm{m~K}$ as the deformation parameter $q$ varies from $1.0$(undeformed) to 
 $1.1$(deformed).

\noindent {{\bf Keywords}: $q$-deformed statistics, blackbody radiation, Rayleigh-Jeans law, Wein's displacement 
law. } 
\end{abstract}

\maketitle

\section{Introduction}

 Statistical mechanics, the art of turning the microscopic laws of physics into a description of 
 nature on a macroscopic scale, has been proved to be a very powerful tools in various domains 
 over the last century. It has been successfully used not only in different branch of 
 physics (e.g. condensed matter physics, high energy physics etc.), but in different areas e.g. 
share price dynamics, traffic control, hydroclimatic fluctuation etc. Results predicted by the 
statistical method are found to be in good agreement with the experiments. The important fact 
is that, without the detailed knowledge of each and every microstate of the concerned system, we 
can predict the macroscopic properties of the system. Things got easier as the connection has 
been build up between the statistical average of the microscopic properties and the macrostates. 
 
 Attempts have been made to generalize this connection mentioned above in recent years
 \cite{Tsallis,Sumiyoshi}. Recently, the techniques of the generalized statistics(popularly known as 
 superstatistics or  $q$-deformed(Tsallis) statistics) have been applied to a large class of complex 
 systems e.g. hydrodynamic turbulence, defect turbulence, share price dynamics, random matrix theory, 
 random networks, wind velocity fluctuations, hydroclimatic fluctuations, the statistics of 
 train departure delays and models of the metastatic cascade in cancerous systems 
 \cite{Tsallis5, Tsallis6, Plastino, Plastino2, Plastino3, Fevzi2, Fevzi3}. 
 In this approach, the key parameter is the inverse temperature parameter $\beta~(=1/k_B T)$ 
 which exhibits fluctuations on a large time scale. One can model these type of complex systems by a 
 kind of superposition of ordinary statistical mechanics with varying temperature 
 parameters, which in short called superstatistics or deformed statistics. The stationary distributions
 of deformed/superstatistical  systems differs from the usual Boltzmann-type statistical mechanics 
 and can exhibit asymptotic power laws or other functional forms in the energy $E$ \cite{cbeck1, cbeck2, cbeck3}.
 
 By using the non-extensive statistical methods we can incorporate the fact of temperature 
 fluctuations and the proceeding sections are devoted mainly for an attempt to give more insight 
 on it\cite{Tsallis2, cohen, Tsallis3}. This approach deals with the fluctuation(deformation) parameter $q$ which corresponds to the degree 
 of the temperature fluctuation effect to the concerned system. In this formalism we can treat 
 our normal Boltzmann-Gibbs statistics as a special case of this generalized one, where 
 temperature fluctuation effects are negligible, corresponds to $q=1.0$(un-deformed statistics). 
 More deviation from $q=1$ denotes a system with more fluctuating temperature \cite{cbeck5, Tsallis4, Atanu}.
 
\section{Temperature fluctuation and the modified entropy}

 The phenomena of temperature fluctuation can be interpretted physically as the deformation of a 
 ideal canonical ensemble to a more realistic case. Ideal canonical ensemble is supposed to be 
 the statistical ensemble that represents the possible states of a mechanical system in thermal
 equilibrium with a heat bath at a fixed temperature (say $T$). Consequently each and every 
 cell(very small identical portions) of the system will be at temperature $T$. To make it more 
 realistic we can think about a modified canonical system which is in thermal equilibrium with 
 a heat bath at a fixed temperature $T$ but there will be a small variation in temperature in 
 different cells(say $T-\delta T$ to $T+\delta T$) though the average temperature of the system 
 will be $T$ still.
 
 A connection between the entropy ($s$) and the number of microstates ($\Omega$) of a system 
 can be derived using intution mainly. We know the entropy is the measure of the number of 
 microscopic configuration, i.e., the degree of randomness of a system. The only thing we can 
 infer clearly is that, they both ($s$ and $\Omega$) will increase (or decrease) together. 
 We can assume $s=f(\Omega)$. Further we know, entropy is additive and the number of microsates
 is multiplicative. These will lead to the form $ s=k_B \ln \Omega $. 
 
 More general assumption can be made which will deform the fundamental connection between $s$ and $\Omega$ as follows
\bea
s=f(\Omega^q)
\eea
This assumption with $q>0$ on the first hand will lead to the deformed/generalized statistics. 
The generalized entropy will take the following form. 
\bea
s_q=k_B \ln_q \Omega
\eea
where the generalized log function is defined as
\bea
\ln_q \Omega = \frac{\Omega^{1-q}-1}{1-q} = x,~ (\rm{say})
\label{eqn:lnq}
\eea
and consequently the generalized exponential function becomes
\bea
e_q^x=\left[1+(1-q) x \right]^\frac{1}{1-q}~\longrightarrow{e^x ~\rm{as}~ q \to 1}
\label{eqn:expq}
\eea
Extremizing $S_q$ subject to suitable constraints yields more general canonical ensembles,
where the probability to observe a microstate with energy $\epsilon_i$ is given by: 
\bea
p_i = \frac{e_q^{- \beta \epsilon_i}}{z} = \frac{1}{z} \left[ 1 - (1-q) \beta \epsilon_i \right]^{\frac{1}{1-q}}
\eea
with partition function $z$ and inverse temperature parameter $\beta=\frac{1}{k_B T}$. 

 This generalized modification can be related to the temperature fluctuation of the system. One 
 can show the connection by expanding the deformed/generalized exponential function as follows:  
\bea
e_q^x &=& \left[1+(1-q) x \right]^{\frac{1}{1-q}} \nonumber \\
&=& 1+x+ q \frac{x^2}{2!}+ q (2q-1) \frac{x^3}{3!}+ q (2q-1) (3q-2) \frac{x^4}{4!}+\cdots
\eea
where, $x=- \beta \epsilon_i= - \frac{\epsilon_i}{k_B T}$.

 The $q$ factors in the expansion can be absorbed in $T$ and those will account for the 
 temperature fluctuation of the system, whereas, $q=1.0$ will lead us back to the normal 
 Boltzmann-Gibbs statistics i.e., the case of zero or negligible temperature fluctuation.
 
 
\section{Effects of $q$-deformed Statistics in Blackbody Radiation}

 In the last section we have seen that we can use the generalized statistical mechanics wherever 
 the system is subjected to the temperature fluctuation. If the temperature 
 fluctuation effect is not negligible enough to disclose itself, then definitely there will be 
 some deviation from the ideal phenomena.
 
 Ideal blackbody radiation (at constant temperature $T$) spectrum, which do not suffer any temperature 
 fluctuation, unfortunately is not  available in nature. If we want to use the blackbody
 radiation formula to determine something in our real world, e.g. the surface temperature of a star from it's 
 spectrum analysis, we can use $q$-deformed statistics and it's impact on the Planck's law of 
 blackbody radiation. The generalized (or $q$-deformed) distribution function takes the 
 following form for small deformation, i.e., small $\mid1-q \mid$(see Appendix, Eq.(\ref{energydensity})) 
  \bea 
f_q = \frac{1}{\left(e_q^{-h \nu/{k_B T}}\right)^{-q}-1} = \frac{1}{\left[1 - (1-q) \frac{h \nu}{k_B T} \right]^{\frac{q}{q-1}} -1}
\eea
Number of photons in the frequency interval $ \nu $ and $ \nu + d \nu $ and volume $V$
\bea
dN = \frac{8 \pi V}{c^3} \cdot \frac{\nu^2 d \nu}{\left[1 - (1-q) \frac{h \nu}{k_B T} \right]^{\frac{q}{q-1}} -1}
\eea
The radiation energy corresponding to photons of frequency lying between $\nu$ and $\nu + d\nu$ is   
\bea
dE (= h\nu \times dN) = \frac{8 \pi h V}{c^3} \cdot \frac{\nu^3 d \nu}{\left[1 - (1-q) \frac{h \nu}{k_B T} \right]^{\frac{q}{q-1}} -1}
\eea

The distribution of the Photon energy density per unit volume is given by(see Appendix, Eq.(\ref{energydensity})) 
\bea
u_q(\nu) d\nu = \frac{8 \pi h}{c^3} \cdot \frac{\nu^3 d\nu}{\left[1 - (1-q) \frac{h \nu}{k_B T} \right]^{\frac{q}{q-1}} -1}
\label{eqed}
\eea
and this is the $q$-deformed blackbody radiation formula.

\begin{figure}[ht!]
\includegraphics[width=11cm]{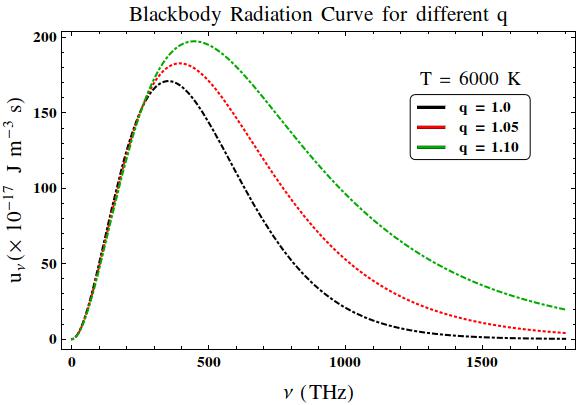}
\caption{\it Spectrum of a Blackbody at temperature $6000~ K$ in $q$-deformed statistics.}
\label{unuq}
\end{figure}

Below in Fig.[\ref{unut}] we have plotted the blackbody radiation curves for different 
temperature without invoking temperature fluctuation i.e. we set $q=1.0$.
\begin{figure}[ht!]
\includegraphics[width=11cm]{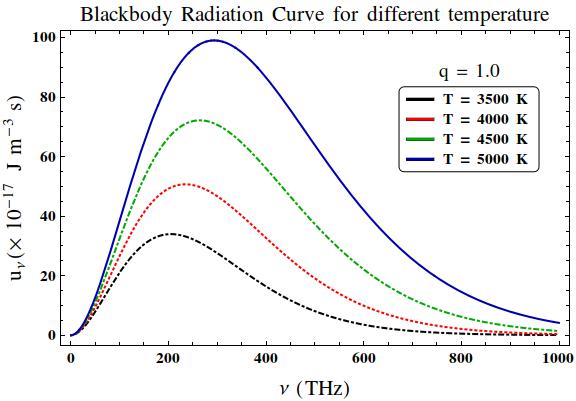}
\caption{\it Spectrum of a Blackbody at different temperatures in undeformed statistics($q=1.0$).}
\label{unut}
\end{figure}
In Fig.[\ref{unuq}] we have plotted the $q$-deformed energy density $u_\nu$ (in units of $J m^{-3} s$ 
(obtained from Eq.(\ref{eqed})) against $\nu$ (in units of Tera Hertz(THz))for different $q$ values (ranging from $q=1$ to $q=1.1$) and constant temperature $ T \sim 6000 K $
(the surface temperature of the sun-like star). 
Also we find that as $q$ increases, not only the location (i.e. $\nu_m$) of the peak of the 
energy distribution $u_\nu$ shifted towards higher $\nu$, the height of the peak also increases(indicates higher rate of radiation). In Table I we have shown the value of the frequency($\nu_m$) at which $u_q(\nu)$ is maximum for different $q$ and also the maximum value of the $q$-generalized energy density.

\begin{center}
\textbf{Table I} \\
\begin{tabular}{ C{3cm} C{3cm} C{4cm} }
\hline\hline
\textbf{$q$} & \textbf{$\nu_m~\rm{(THz)}$} & \textbf{$u_q(\nu)_{max}~\rm{(J~m^{-3}~s)} $} \\
\hline

$~1.0~$ & $~ 352 ~$ & $~ 171 \times 10^{-17}~$ \\

$~1.05~$ & $~ 395~ $ & $~ 183 \times 10^{-17}~$ \\

$~1.10~$ & $ ~446~ $ & $ ~198 \times 10^{-17}~$ \\
\hline\hline
\end{tabular}
\end{center}

The radiation curves corresponding to different $q$ values for a given $T$ appears to be 
different radiation curves corresponding to different $T$ values for $q=1$ (undeformed case).


\subsection{Generalized ($q$-deformed) Wien's Law and Rayleigh-Jeans Law}
\noindent We next study $q$-deformed blackbody radiation formula and it's different limts.

\noindent{\bf{$q$-deformed Wein's law}}:~In the limit of low temperature $T$ or high frequency $\nu$ 
 (i.e. $ x= \frac{h \nu}{k_B T} \gg 1$), we find 
\bea
u_q(\nu) = \frac{8 \pi h}{c^3} \frac{\nu^3}{\left[1 - (1-q) \frac{h \nu}{k_B T} \right]^{\frac{q}{q-1}}}
\label{eqw}
\eea
This is the generalized Wien's law, which reduces the usual(undeformed) Wein's law 
\bea
u(\nu) = \frac{8 \pi h}{c^3} \nu^3 ~e^{-h\nu/k_B T} 
\label{eqw1}
\eea
in the limit $q \to 1$. 

\noindent{\bf{$q$-deformed Rayleigh-Jeans Law}}:~In the limit of high temperature $T$ or low frequency $\nu$ 
 (i.e. $ x= \frac{h \nu}{k_B T} \ll 1$), we find
\bea 
u_q(\nu) = \frac{8 \pi (k_B T)^3}{h^2 c^3} \cdot \frac{x^3}{q x+\frac{q}{2!}x^2+ \frac{q(2-q)}{3!} x^3 + \cdots } 
\label{eqrj}
\eea
which is the generalized Rayleigh-Jeans law. This reduces to the usual(undeformed) Rayleigh-Jeans law
\bea
u(\nu) = \frac{8 \pi k_B T \nu^2 }{c^3}  
\label{eqrj1}
\eea
in the limit $q \to 1$.  

\begin{figure}[ht!]
\includegraphics[width=11cm]{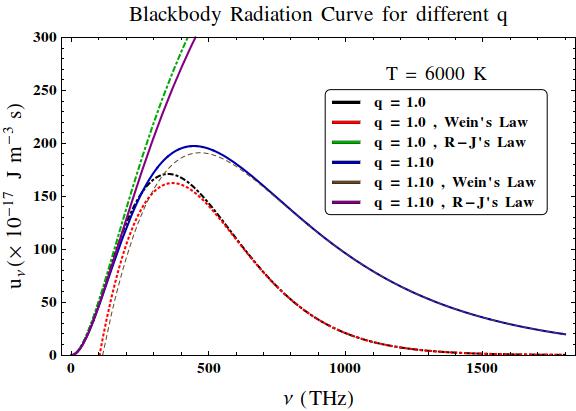}
\caption{\it Spectrum of a Blackbody along with the Wien's Law and Rayleigh-Jeans Law at temperature $6000~ K$ in $q$-deformed statistics.}
\label{uni}
\end{figure}

In Fig.[\ref{uni}], we have shown the generalized Wien's law and Rayleigh-Jeans Law for 
different amount of $q$-deformation using Eqs.(\ref{eqw}) and (\eqref{eqrj}) respectively 
along with the blackbody radiation curves obtained from Eq.(\ref{eqed}). 
 
Plots corresponding to the undeformed cases ($q=1$) are also shown in the same plot. 
We find that as $q$ varies from $1.0$ to $1.1$, plots are getting well seperated at high and low frequency 
region, respectively. Also in addition, for the Planck's blackbody radiation formula, the location 
of the peak (at which the radiation density is maximum) gets shifted towards high frequency value 
as $q$ increases from $1.0$ to $1.1$.

\subsection{Generalized Stefan's law of radiation}

By integrating the frequencies for all values, we find the energy emitted by the blackbody 
per unit time (i.e. power) per unit surface area as 
\bea
 P_q = \int_0^{\infty} d \nu u_q(\nu) = \frac{8 \pi k_B^4}{h^3 c^3} \cdot T^4 \cdot \int_0^{\infty} \frac{x^3~dx}{\left[1 - (1-q) x \right]^{\frac{q}{q-1}} -1} = \frac{4}{c} \cdot \sigma_q ~T^4  
\eea 
This is the generalized or $q$-deformed form of the Stefan's law of blackbody radiation. Here 
$\sigma_q$ is the $q$-deformed Stefan's constant. Note that $\sigma_{q=1.05} \sim 7.6 \times 10^{-8}~\rm{J~sec^{-1} m^{-2} K^{-4}}$, 
$\sigma_{q=1.1} \sim 1.1 \times 10^{-7}~\rm{J~sec^{-1} m^{-2} K^{-4}}$ in deformed scenario. In the 
undeformed scenario, it becomes $\sigma_{q=1.0} \sim 5.7 \times 10^{-8}~\rm{J~ sec^{-1} m^{-2} K^{-4}}$. 

\subsection{Generalized Wien's Displacement Law}
 
 In the $q$-deformed formalism, the photon energy density per unit volume is given by 
\bea
u_q(\lambda) = \frac{8 \pi h c}{\lambda^5} \cdot \frac{1}{\left[1 - (1-q) \frac{h c}{\lambda k_B T} \right]^{\frac{q}{q-1}} -1} 
\label{eqn1}
\eea
Wien's displacement law states that the black body radiation curve for different temperatures peaks at a wavelength inversely proportional to the temperature. Let
$ \lambda_m $ be the wavelength at which $u_q(\lambda)$ is maximum at temperature $T$, i.e.
\bea
\left[\frac{du_q(\lambda)}{d \lambda} \right]_{\lambda=\lambda_m} = 0
\label{eqn2}
\eea
Using Eqs.(\ref{eqn1}) and (\ref{eqn2}) we obtain
\bea
\frac{8 \pi h c}{\lambda_m^6}\frac{1}{\left(e_q^{-hc/\lambda_m k_B T}\right)^{-q}-1} \left\lbrace \frac{hc}{\lambda_m k_B T} \frac{q \left(e_q^{-hc/\lambda_m k_B T}\right)^{-1}}{\left(e_q^{-hc/\lambda_m k_B T}\right)^{-q}-1}-5\right\rbrace =0
\label{eqn3}
\eea
This equation(Eq.(\ref{eqn3})) leads to
\bea
\frac{y}{5}+\frac{1}{q} e_q^{-y} - \frac{1}{q} \left( e_q^{-y} \right)^{1-q} = 0 ~~\rm{with}~~ y=\frac{h c }{\lambda_m k_B T}
\label{eqnd}
\eea
which is the $q$-deformed version of Wein's displacement law. In the limit $ q \rightarrow 1.0 $ 
(no temperature fluctuation) it becomes 
\bea 
\frac{y}{5}+e^{-y}-1 = 0 
\label{eqnud}
\eea
which is the normal (undeformed) version of Wein's displacement law.
Below in Fig.[\ref{lmt}] we have shown the variation of $ \lambda_m T $ with the deformation parameter 
$q$. Clearly, the value of $\lambda_m T$ is found to be decreasing with increasing $q$.  This will result in a shift of the peak of the blackbody spectrum in the direction of 
 decreasing wavelength $\lambda$(increasing frequency $\nu$) for a fixed temperature.
\begin{figure}[ht!]
\centering
\begin{minipage}{0.45\textwidth}
\includegraphics[width=8.0cm]{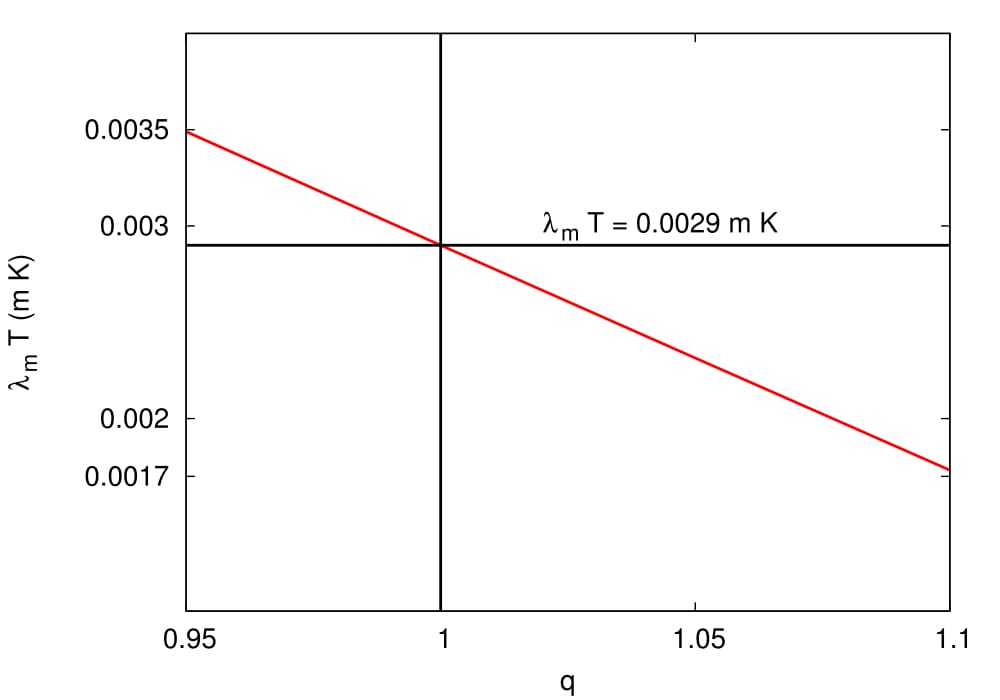} 
\caption{\it The variation of $ \lambda_m T $ with $q$}
\label{lmt}
\end{minipage}
\begin{minipage}{0.5\textwidth}
\includegraphics[width=8.0cm]{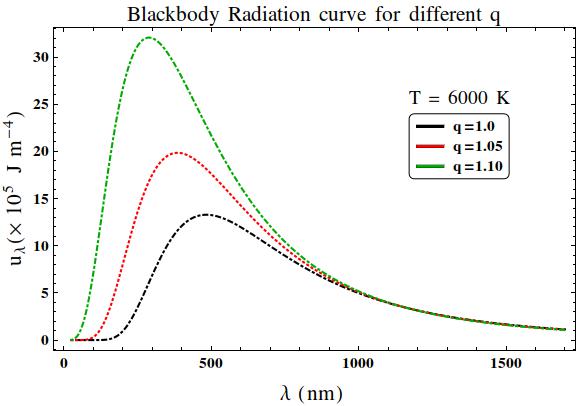}
\caption{\it $q$-deformed blackbody spectrum (at temperature $6000~ K$).}
\label{ulam}
\end{minipage}
\end{figure}
In Fig.[\ref{ulam}] we have plotted the blackbody radiation energy density as a function of the wavelength
corresponding to the fixed temperature $6000~K$ for different $q$. Shift in $\lambda_m$ clearly can be 
seen from the plot. In Table II, we have shown the solution of Eq.[\ref{eqnd}] and the corresponding 
$ \lambda_m T $ for different $q$ values.  We see that as $q$ increases the value of $\lambda_m$
(corresponding to the peak energy density) decreases, while the height of the peak increases. 
\begin{center}
\textbf{Table II} \\ 
\begin{tabular}{ C{3cm} C{3cm} C{3cm} }
\hline\hline
\textbf{$q$} & \textbf{$y$} & \textbf{$ \lambda_m T~\rm{(m~ K)} $} \\
\hline

$~0.95~$ & $~ 4.126 ~$ & $~ 3.5 \times 10^{-3}~$ \\

$~1.0~$ & $~ 4.965 ~$ & $~ 2.9 \times 10^{-3}~$ \\

$~1.05~$ & $~ 6.156~ $ & $~ 2.3 \times 10^{-3}~$ \\

$~1.10~$ & $ ~8.143~ $ & $ ~1.7 \times 10^{-3}~$ \\
\hline\hline
\end{tabular}
\end{center}



\section{Conclusion}
We study the Planck's law of blackbody radiation and it's high frequency limit(Wein's law) and 
low frequency limit(Rayleigh-Jeans law) from the point of view $q$-deformed(Tsallis).
The radiation curves corresponding to different $q$ values for a given $T$ appears to be 
different radiation curves corresponding to different $T$ values for $q=1$ (undeformed case). 
The location i.e. $\nu_m$ of the peak of the energy distribution $u_\nu$ (corresponding to different $q$ ) shifted towards higher $\nu$. 
In the case of Wein's law, we find that $\lambda_m T$ varies from $0.0029~\rm{m~K}$ to 
$0.0017~\rm{m~K}$ as the deformation parameter $q$ varies from $1.0$ to  $1.1$. 
 

\section*{ACKNOWLEDGMENTS}
\noindent The authors would like to thank Selvaganapathy J. for discussions. 
\section*{APPENDIX}
\subsection*{Indicial properties of $q$-deformed exponential function for small deformation}

 From Eq.(\ref{eqn:expq}), keeping only first order in $(1-q)$,
\bea
e_q^a \cdot e_q^b &=& \left[1+(1-q) a \right]^\frac{1}{1-q} \cdot \left[1+(1-q) b \right]^\frac{1}{1-q} \nonumber \\
&=& \left[1+ (1-q) (a+b) + (1-q)^2 ab \right]^\frac{1}{1-q} \nonumber \\
& \simeq & e_q^{a+b} 
\label{smlmul}
\eea

 Similarly, neglecting higher order terms we get(for small $\mid1-q\mid$),
\bea
\left(e_q^a \right)^b &=& \left[1+(1-q) a \right]^\frac{b}{1-q} \nonumber \\
&=& \left[1+ (1-q) ab + \frac{b(b-1)}{2!}(1-q)^2 a^2+ \cdots \right]^\frac{1}{1-q} \nonumber \\
& \approx & e_q^{ab} 
\label{smlpwr}
\eea

 The energy density of the oscillators having frequency in the range $\nu$ to $\nu+d\nu$ 
\bea
u_q(\nu)=\frac{g(\nu)}{V} \left\langle E(\nu) \right\rangle = \frac{8 \pi \nu^2}{c^3} \left\langle E(\nu) \right\rangle
\eea

Now the average energy per oscillator is defined as(in $q$-deformed scenario) \cite{Tsallis, guo, Swamy}
\bea
\left\langle E(\nu) \right\rangle = \frac{\sum_i \epsilon_i p_i^q}{\sum_i p_i^q}
\eea
where the factor $\sum_i p_i^q$ stands for the normalization. Writing $x=\frac{h \nu}{k_B T}$ we get
\bea
\left\langle E(\nu) \right\rangle = h \nu \frac{\sum_{n=0}^{\infty} n \left(e_q^{-n x}\right)^q}{\sum_{n=0}^{\infty} \left(e_q^{-n x}\right)^q}
\label{avgenrgy}
\eea

Now in the small deformation approximation(i.e., small $ \mid 1-q \mid $) using Eq.(\ref{smlpwr})
\bea
\left(e_q^{-n x}\right)^q \approx \left(e_q^{-x}\right)^{q n}
\eea
Putting this back to the Eq.(\ref{avgenrgy}) we get
\bea
\left\langle E(\nu) \right\rangle = h \nu \frac{\sum_{n=0}^{\infty} n \left(e_q^{-x}\right)^{q n}}{\sum_{n=0}^{\infty} \left(e_q^{-x}\right)^{q n}}
= \frac{h \nu}{\left(e_q^{-x}\right)^{-q}-1}
\eea
where we used the fact that,
\bea
\sum_{0}^{\infty} n y^n = \frac{y}{\left(1-y\right)^2}~\rm{and}~ \sum_{0}^{\infty} y^n = \frac{1}{1-y}
\eea
This will give the following compact form for the $q$-deformed version of the energy density
\bea
u_q(\nu)= \frac{8 \pi h \nu^3}{c^3} \frac{1}{\left(e_q^{-x}\right)^{-q}-1}
\label{energydensity}
\eea

\end{document}